\documentstyle[epsf,amstex,amssymb,righttag,12pt]{article}
\newtheorem{th}{Theorem}
\newtheorem{pro}{Proposition}

\begin{document}
\title{\Large\sc Darboux transformation for the \\
Manin-Radul supersymmetric KdV equation}
\author{Q. P. Liu\thanks{On leave of absence from
Beijing Graduate School, CUMT, Beijing 100083, China}
\thanks{Supported by {\em Beca para estancias temporales
de doctores y tecn\'ologos extranjeros en
Espa\~na: SB95-A01722297}}
   $\,$ and M. Ma\~nas\thanks{Partially supported by CICYT:
 proyecto PB92-019}\\
 Departamento de F\'\i sica Te\'orica, Universidad Complutense\\ 
E28040-Madrid, Spain.}

\date{}

\maketitle

\begin{abstract}
In this paper we present a vectorial  Darboux transformation,
in terms of ordinary determinants,
for the supersymmetric extension of the
Korteweg-de Vries equation proposed by Manin and Radul.
It is shown how this transformation reduces to the
Korteweg-de Vries equation.
Soliton type solutions are constructed by dressing the vacuum 
and we present some relevant plots.  
\end{abstract}
\newpage

\section{Introduction}

The Korteweg-de Vries (KdV) equation was embedded in a  
supersymmetric framework for the first time by
Manin and Radul in \cite{mr}. Since then a number of 
integrable equations have been extended in this way.
The role of the KdV equation and
its Virasoro constraints in two dimensional quantum gravity
\cite{bar,ds,gm} lead the group of Alvarez-Gaum\'e to search
for analogous structures for supersymmetric two dimensional
quantum gravity \cite{ag1,ag2}. In turn, this motivated
the study of Virasoro constraints for the supersymmetric
Kadomtsev-Petviashvilii (KP) hierarchies avaliable \cite{mmm},
which is connected with the study of additional symmetries
of these hierarchies \cite{mmm,das,ss}. These results indicated
that the supersymmetric extensions of the KdV
equation, in particular the Manin-Radul super KdV (MRSKdV),
might be relevant in the study of susy 2d quantum gravity.

The search of solutions of the Manin-Radul super KP started
with the work of of Radul \cite{r} on algebro-geometric
 type solutions.
Then in \cite{ueno}, from a Sato Grassmannian approach,
the construction of solutions was outlined, nevertheless
one can not find explicit examples in this paper. Recently
in \cite{Ibt} some explicit solutions were obtained.

The MRSKdV system is defined in terms of three independent variables
$\vartheta,x,t$, where $\vartheta\in\Bbb C_{\text{a}}$ is an odd supernumber,
and $x,t\in\Bbb C_{\text{c}}$ are even supernumbers, and 
two dependent variables $\alpha(\vartheta,x,t), 
u(\vartheta,x,t)$, where $\alpha$ is an odd function taking values
in $\Bbb C_{\text{a}}$ and $u$ is even function with values in
$\Bbb C_{\text{c}}$.
A basic ingredient is a superderivation  
defined by $D:=\partial_{\vartheta}+\vartheta \partial_x$.
The system is
\begin{equation}\label{mr}
\begin{aligned}
 \alpha_t&={1\over 4} (\alpha_{xxx}+3(\alpha D\alpha)_x+6(\alpha u)_x),\\
 u_t&={1\over 4}(u_{xxx}+6uu_x+3\alpha_xDu+3\alpha (Du_x)),
\end{aligned}
\end{equation}
where we use the notation 
$f_x:=\partial f/\partial x$ and $f_t:=\partial f/\partial t$.


The following linear system for the wave function $\psi(\vartheta, x,t)$,
that takes values in the Grassmann algebra $\boldsymbol\Lambda=\Bbb
C_{\text{c}}\oplus\Bbb C_{\text{a}}$,
\begin{equation}\label{linear}
\begin{aligned} 
\psi_{xx}+\alpha D\psi +u\psi-\lambda \psi&=0,\\
\psi_t-{1\over 2}\alpha(D\psi)_x-\lambda\psi_x-{1\over2}u\psi_x
+{1\over 4}\alpha_x D\psi+{1\over 4}u_x\psi&=0,
\end{aligned}
\end{equation}
where the spectral parameter $\lambda\in\Bbb C_{\text{c}}$ is an even
supernumber, has as its compatibility condition Eqs. (\ref{mr}), and therefore
it can be considered as a Lax pair for it.

Our aim in this paper is to extend a well known tool, Darboux
transformations, in integrable
system theory to the supersymmetric case. This tool is
a well established  scheme  in dealing with integrable equations
and its solutions \cite{ms}. Given an integrable equation and its Lax pair
 the Darboux technique consists
of transforming simultaneously both fields and wave functions.
For the KdV equation the Lax pair is essentially the Schr\"odinger equation,
and this was precisely the equation where Darboux developed his
technique. On the one hand, recently
 one of the authors extended the 
standard Darboux transformations to the supersymmetric KdV, \cite{liu}.
On the other hand,  the other author 
  has been involved recently in generalizing the standard Darboux techniques
to a vectorial Darboux transformation, \cite{fm,mm}.

In this paper we  present a vectorial Darboux transformation
for the MRSKdV, this transformation is represented in terms
of ordinary determinants of an even operator. Nevertheless, we do
obtain essentially supersymmetric solutions. Indeed any
arbitrary solution  can be used as seed solution to dress, and obtain
therefore large families of new solutions having the seed solution
as background.

The layout of the paper is as follows. 
In \S 2 we include the main results
of the paper, namely the vectorial Darboux transformation 
for the MRSKdV equation
(\ref{mr}). There, we also consider the reduction to the KdV equation.
Next, in \S 3 we study some explicit solutions selected among the large
classes of explicit solutions offered by this method. In particular,
we dress the vacuum solution to obtain soliton type solutions.
Here we also give some plots showing the behaviour of the field $\alpha$
(as the ones for the $u$ are similar we do not include them here). 
We end with some conclusions and remarks in section 4.

\section{Vectorial Darboux transformation}
The linear system (\ref{linear}) is of a scalar nature,
$\lambda\in\Bbb C_{\text{c}}$,
$\psi(\vartheta,x,t)\in\boldsymbol\Lambda$.
 Nevertheless, it is possible to give a vector extension
of these linear problem.
Indeed, we may replace $\boldsymbol \Lambda$ 
by an arbitrary linear Grassmann space 
$\cal E$ over $\boldsymbol \Lambda$ 
and take $b$ as an $\cal E$-valued eigenfunction, then the spectral parameter
can be taken as  $L\in  \;\text{L}(\cal E_{\underline{0}})\oplus
\text{L}(\cal E_{\underline{1}})$,  an even operator.

Namely, the linear system
\begin{equation}\label{veclin}
\begin{aligned}
 b_{xx}+\alpha Db+ub-Lb&=0,\\
 b_t-{1\over2} \alpha (Db_x) -Lb_x-{1\over 2}ub_x+
{1\over 4}\alpha_x Db
+{1\over 4}u_xb&=0,
\end{aligned}
\end{equation}
has as its compatibility condition the
MRSKdV system (\ref{mr}). 

Notice  that Eqs. (\ref{mr}) is also the compatibility
condition of adjoint linear system:
\begin{equation}\label{aveclin}
\begin{aligned}
\beta_{xx}+D(\alpha \beta )+u\beta -\beta M&=0,\\
\beta_t+{1\over 2}\alpha D\beta_x-\beta_xM-{1\over 2}
(u+D\alpha )\beta_x +{1\over 4}D(\alpha_x\beta)+{1\over 4}u_x\beta&=0,
\end{aligned}
\end{equation}
where $\beta(\vartheta,x,t)\in\cal E^*$ is a linear function on the
supervector space $\cal E$, and 
$M \in \text{L}(\cal E_{\underline{0}})\oplus\text{L}(\cal E_{\underline{1}})$.

In order to construct Darboux transformation for these linear systems 
we need to introduce an even operator, say $V$, to this end
we assume that $b(\vartheta, x,t)\in\cal E_{\underline{0}}$
is an even vector
and   $\beta\in\cal E_{\underline{1}}^*$ an odd functional.

\begin{pro}
 Let $b(\vartheta,x,t)$ and $\beta(\vartheta,x,t)$
satisfy Eqs. (\ref{veclin}) and (\ref{aveclin}), respectively.
Then, there exists
a potential operator $V(\vartheta, x,t)\in
\text{L}(\cal E_{\underline{0}})\oplus
\text{L}(\cal E_{\underline{1}})$ given by
\begin{equation}\label{pont}
\begin{aligned}
DV=&b\otimes \beta, \\
V_t=&LV_x +V_xM -D(b_x\otimes\beta_x+{1\over 2}uDV)
-{1\over 4}\alpha_x DV\\ 
&-{1\over2}(Db)\otimes((D\alpha)\beta-\alpha
(D\beta))+{1\over2}\alpha(b\otimes \beta_x -b_x\otimes \beta )
\end{aligned}
\end{equation}
such that
\begin{equation}\label{constraint}
LV-VM =D(b_x\otimes\beta-b\otimes \beta_x)-
\alpha b\otimes \beta.
\end{equation}
\end{pro}

{\bf Proof}: A direct calculation shows that $DV_t=(DV)_t$ holds.
We proceed by checking that the identity
$$
D\left(LV-VM-D(b_x\otimes\beta-b\otimes\beta_x)+\alpha b\otimes\beta\right)
=0
$$ 
holds. A tedious but straightforward calculation shows
$$
\partial_t\left(LV-VM-D(b_x\otimes\beta-b\otimes\beta_x)+\alpha b\otimes\beta\right)
=0.\Box
$$

Now, we state the  main result of the paper. 

\begin{th}
Let $b(\vartheta,x,t)\in\cal E_{\underline{0}}$ be an even
vector satisfying Eq. (\ref{veclin}), 
$\beta(\vartheta,x,t)\in\cal E_{\underline{0}}^*$  an odd functional
solving Eq. (\ref{aveclin}) and $V\in \text{L}(\cal
E_{\underline{0}})\oplus\text{L}(\cal E_{\underline{1}})$  a non
singular even
operator, $\det V_{\text{body}}\neq 0$, defined in terms
of the compatible Eqs.
(\ref{pont}) and (\ref{constraint}).
Then, the objects
\begin{align*}
&\hat{b}:=V^{-1}b,\; \hat{\beta}:=\beta V^{-1},\;
 {\hat{L}}:=M,\; {\hat{M}}:=L\\
&\hat\alpha=\alpha-2D^3\ln\det V,\\
&\hat u=u+2\hat\alpha D\ln\det V+2\left(
\frac{\sum_{j}D\beta_j\;\det V_j}{\det V}\right)_x,
\end{align*}
where  $V_j$ is an operator with associated supermatrix
obtained from the corresponding one of $V$ by replacing the $j$-th
column by $b$, satisfy the Eqs. (\ref{veclin}) and (\ref{aveclin}) whenever
the unhatted variables do. Thus, $\hat\alpha$ and $\hat u$ are
new solutions of the MRSKdV (\ref{mr}).
\end{th}
{\bf Proof:}
Is a tedious but straightforward calculation to check that
\begin{align*}
&\hat{b}=V^{-1}b,\; \hat{\beta}=\beta V^{-1},\;
 {\hat{L}}=M,\; {\hat{M}}=L\\
&\hat{\alpha}=\alpha-2\langle\beta,\; V^{-1}b \rangle_x,\\
&\hat{u}=u+2\langle D\beta, V^{-1}b\rangle_x+2\alpha \langle
\beta,V^{-1}b\rangle+2\langle\beta,V^{-1}b\rangle
\langle\beta,V^{-1}b\rangle_x,
\end{align*}
satisfy Eqs. (\ref{veclin}) and (\ref{aveclin}).
Observe that form $DV=b\otimes\beta$  one has the relation
$\text{Tr}(DV\cdot V^{-1})=\langle\beta, V^{-1} b\rangle$,
and therefore 
\[
\langle\beta, V^{-1}b\rangle=D\ln\det V,
\]
where we are using standard traces and determinants.
Notice also that, using Cramer's rule, we have
\[
\langle D\beta,V^{-1}b\rangle=\frac{\sum_{j}D\beta_j\;\det V_j}{\det V}.
\]

These three remarks  lead to the desired result. $\Box$

\paragraph{Reduction to KdV}
The MRSKdV system (\ref{mr}) reduces to the KdV equation when $\alpha=0$.
Our Darboux transformation is in fact compatible with this
reduction, giving in this manner Darboux transformation for the
KdV equation. To see this let us first note 
that with the splitting $b(\vartheta,x,t)=b_0(x,t)+
\vartheta b_1(x,t)$ ,  $\beta(\vartheta,x,t)=\beta_1(x,t)+
\vartheta\beta_0(x,t)$ and $V(\vartheta,x,t)=V_0(x,t)+
\vartheta V_1(x,t)$, Eq. (\ref{pont}) reads
\[
V_1=b_0\otimes \beta_1, \qquad V_{0x}=b_1\otimes\beta_1+
b_0\otimes\beta_0
\]

The linear systems (\ref{veclin}) and (\ref{aveclin})
 with $\alpha=0$ are
\[
b_{xx}+ub=Lb, \qquad b_t=Lb_x+{1\over 2}ub_x-{1\over 4}u_xb
\]
and
\[
\beta_{xx}+u\beta=\beta M, \qquad 
\beta_t=\beta_x M+{1\over 2}u\beta_x-{1\over 4}u_x\beta
\]

To proceed further, we assume $\beta_1=b_1=0$ so that $V_1=0$. Then,
our potential satisfies
\[
V_{0,x}=b_0\otimes \beta_0,
\]
In this case, our Darboux transformation tells us
\begin{align*}
\hat{\alpha}&=-2\vartheta\langle\beta_0, V_{0}^{-1}b_0\rangle_x,\\
{\hat{u}}&=u+2\langle\beta_0,V_{0}^{-1}b_0\rangle_x
\end{align*}
 and since ${\hat\alpha}D\hat{b}=0$ we conclude that 
$\hat{u}$ satisfies the equations
\begin{align*}
\hat{b}_{xx}+\hat{u}\hat{b}-\hat{L}\hat{b}&=0, \\
\hat{b}_t-\hat{L}\hat{b}_x-{1\over 2}\hat{u}\hat{b}_x+
{1\over 4}\hat{u}_x\hat{b}&=0,
\end{align*}
and therefore is a new solution of the KdV equation.
Notice that here we have a subtle point,
observe that $\hat\alpha$ is not zero but in turn does not
appear in the evolution equations because its particular structure.

Observe also that using the formula 
$\text{Tr}(V_{0,x}\cdot V_0^{-1})=(\ln\det V_0)_x$,
the transformation for field u can be rewritten neatly as 
$$
{\hat{u}}=u+2(\ln\det V_0)_{xx},
$$
which a standard form for the solutions of the KdV equation \cite{an}.

\section{Exact solutions of the MRSKdV}
 
Among the large classes of solutions provided by
the just presented  vectorial Darboux transformation,
 in this section we select
some relevant example  by dressing  the vacuum
solution $\alpha=0$, $u=0$. In doing so we obtain solutions,
that for simplicity we denote by $\alpha, u$ erasing the
hat, of the MRSKdV equation (\ref{mr}) which can be considered
as a superextension of the soliton solutions of KdV.

Inserting $\alpha=u=0$ in  the linear systems
one gets the equations for $b$
\begin{align*}
&\begin{cases}
b_{xx}=Lb, \\ b_t=Lb_x,
\end{cases}\\
&\begin{cases}
\beta_{xx}=\beta  M,\\
\beta_t=\beta_x M
\end{cases}
\end{align*}

For simplicity we take $L$, $ M$ as diagonal even matrices,
 $L=\text{diag}(\ell_{1}^{2},\cdots,\ell_{n}^{2}) $ and $ M=
\text{diag}( m_{1}^2,\cdots,  m_{n}^{2})$, $\ell_j,m_j\in\Bbb
C_{\text{c}}$, $j=1,\dots, n$.
Then, the functions $b$ and $\beta$ have the following form:
\[
 b_i=c_{i,+}\exp(\eta_{i}) +c_{i,-}\exp(-\eta_{i}),\qquad
 \beta_j =\kappa_{j,+}\exp({\xi}_j)+
\kappa_{j,-}\exp({-\xi}_j)
\]
where $\eta_i(x,t):=\ell_i(x+\ell_{i}^{2}t)$ and 
$\xi_j(x,t):= m_j(x+ m_{j}^{2}t)$.

The operator $V$, whenever 
$(\ell_{i}^2- m_{j}^{2})_{\text{body}}\neq 0$ for all $i,j$,
is determined by the constraint (\ref{constraint}), namely
\[
V_{ij}={1\over \ell_{i}^2- m_{j}^{2}} D\varphi_{ij}
\]
where 
\begin{align*}
\varphi_{ij}=&-\ell_i(c_{i,-}\exp(-\eta_i)-c_{i,+}\exp(\eta_i))
(\kappa_{j,-}\exp(-\xi_j)+\kappa_{j,+}\exp(\xi_j))\\ 
&+ m_j(c_{i,-}\exp(\eta_i)+c_{i,+}\exp(\eta_i))
(\kappa_{j,-}\exp(\xi_j)-\kappa_{j,+}\exp(\xi_j))
\end{align*}
and $c_{i,\pm}=c_{i,\pm}(\vartheta)$ and 
$\kappa_{j,\pm}=\kappa_{j,\pm}(\vartheta)$.

The expression can be made explicitly by means of substitutions
$$
c_{i,\pm}=c_{i,\pm}^0+\vartheta c_{i,\pm}^1, \qquad
\kappa_{j,\pm}=\kappa_{j,\pm}^1+\vartheta\kappa_{j,\pm}^0,
$$
where the superfix indicate the parities of the variables.

Indeed, we have
$$
V_{ij}=V_{ij}^{0}+\vartheta V_{ij}^{1}
$$
with
\begin{alignat*}{3}
V_{ij}^{0}=& {1\over \ell_i+ m_j}
&&\hspace*{-.5cm}\big((c_{i,+}^{0}\kappa_{j,+}^{0}+
c_{i,+}^{1}\kappa_{j,+}^{1})\exp(\eta_i+\xi_j)\\&&&-
(c_{i,-}^{1}\kappa_{j,-}^{1}
+c_{i,-}^{0}\kappa_{j,-}^{0})\exp(-\eta_i-\xi_j)\big)\\
&+{1\over \ell_i- m_j}&&
\big((c_{i,+}^{1}\kappa_{j,-}^{1}
+c_{i,+}^{0}\kappa_{j,-}^{0})\exp(\eta_i-\xi_j)\\
&&&-(c_{i,-}^{1}\kappa_{j,+}^{1}
+c_{i,-}^{0}\kappa_{j,+}^{0})\exp(-(\eta_i-\xi_j))\big)
\end{alignat*}
and 
\begin{align*}
V_{ij}^{1}=&c_{i,+}^{0}\kappa_{j,+}^{1}\exp((\eta_i+\xi_j))
+c_{i,-}^{0}\kappa_{j,-}^{1}\exp(-\eta_i-\xi_j)\\&+
c_{i,-}^{0}\kappa_{j,+}^{1}\exp(-(\eta_i-\xi_j))+
c_{i,+}^{0}\kappa_{j,-}^{1}\exp(\eta_i-\xi_j)
\end{align*}

As a particular example one can pick $b_i$  as above 
and $\beta_j=\kappa_{j}^{1}+\vartheta\kappa_{j}^{0}$, which 
requires $ m_j=0$.
In this case, the potential is
\begin{align*}
V_{ij}={1\over \ell_i}
&(\kappa_{j}^{0}c_{i,+}^{0}-\kappa_{j}^{1}
c_{i,+}^{1})\exp(\eta_j)+(\kappa_{j}^{1}c_{i,-}^{1}-
\kappa_{j}^{0}c_{i,-}^{0})\exp(-\eta_i)\\
&+
\vartheta \ell_i\kappa_{j}^{1}(c_{i,+}^{0}\exp(\eta_{i})+
c_{i,-}^{0}\exp(-\eta_{i}))
\end{align*}

To illustrate this family of solutions we shall consider
the simplest case of a one dimensional space $\cal E$ 
(for simplicity we omit the index 1 next). 
Hence, we have
\begin{align*}
V=&{1\over \ell}((\kappa^{0}c_{+}^{0}-\kappa^1c_{+}^{1})\exp(\eta)+
(\kappa^{1}c_{-}^{1}-\kappa^0c_{-}^{0})\exp(-\eta)\\ &+
\vartheta \ell\kappa^1
(c_{+}^{0}\exp(\eta)+c_{-}^{0}\exp(-\eta))
\end{align*}

To invert the operator $V$, we have to separate 
its body from its soul. 
We choose the supernumbers 
$\kappa^0,\ell,c_{\pm}^0\in\Bbb C$ to have vanishing soul,
so  that the potential operator $V$ takes the following 
form:
$$
V={1\over \ell}(V_{\text{body}}+V_{\text{soul}})
$$
where 
$ V_{\text{body}}:=\kappa^0(c_{+}^{0}\exp(\eta)-
c_{-}^{0}\exp(-\eta))$
and
$ V_{\text{soul}}:=\kappa^1(-c_{+}^{1}\exp(\eta)+
c_{-}^{1}\exp(-\eta)-\vartheta \ell b_0)$, 
 we notice that $b=b_0+\vartheta b_1$ with
 $b_0=c_{+}^{0}\exp(\eta)+c_{-}^{0}\exp(-\eta)$ and
$b_1=c_{+}^{1}\exp(\eta)+c_{-}^{1}\exp(-\eta)$.

The constitutive elements for our solution are now 
\begin{align*}
\langle\beta, V^{-1}b\rangle=& \ell\beta{b_0\over V_{\text{body}}}-
2\vartheta{\ell\kappa^0\kappa^{1}(c_{+}^{0}c_{-}^{1}-c_{-}^{0}c_{+}^{1})
\over (V_{\text{body}})^{2}},\\
\langle D\beta, V^{-1}b\rangle=&
\kappa^0 \ell{b\over V_{\text{body}}}-
{\ell\kappa^0\kappa^1\over (V_{\text{body}})^{2}}
(- c_{+}^{1}c_{+}^{0}\exp(2\eta)+
c_{-}^{1}c_{-}^{0}\exp(-2\eta)
\\ &-c_{+}^{1}c_{-}^{0}+c_{-}^{1}c_{+}^{0})
+{\vartheta \ell\kappa^0\kappa^1\over (V_{\text{body}})^{2}}(2c_{+}^{1}c_{-}^{1}- \ell b_{0}^{2})
\end{align*}

It is easy to see that this solution can be considered as a supersymmetric
extension of the soliton solution of the KdV equation.
In fact, the constants can be chosen in such a way that the functions
depending solely on $x,t$ appearing as multiplicative coefficients
are exponentially localized in $x$ and travel with
constant speed. Moreover, the KdV soliton solution
appears as a particular coefficient, $(b_0/V_{\text{body}})_x$.
Thus, our solution  
can be considered as a supersymmetric
deformation of the standard KdV soliton.

We write $\alpha(\vartheta,x,t)=\beta(\vartheta) f(x,t)-\vartheta
\kappa^1\kappa^0g(x,t)$, and we plot the functions $f$ and $g$.
The function $f$, which is plotted in Figure 1,
 is just a KdV 1-soliton solution while $g$, plotted in Figure 2, is a
exponentially localized regular solution that travels with constant
speed, now its shape is more involved that in the KdV soliton. 

\section{Conclusions and remarks}

We have constructed a Darboux transformation of a vector nature
for the Manin-Radul supersymmetric KdV system. We have further shown
that our Darboux transformation can be reduced to the KdV equation and
is very effective when  exact solutions are needed.
The vectorial Darboux transformation is given in terms of solutions 
of the Lax pair and its adjoint which have a well defined and opposite
parities. This implies that the Darboux operator is even and that the
new solution can be expressed in terms of ordinary determinants.
However, this absence of superdeterminants is not a drawback
because the proposed technique gives an efficient method of
construction of genuine supersymmetric solutions.

We further remark that
the basic Darboux transformations considered in \cite{liu} can be
iterated so that the Crum type transformations may
be obtained. In this case, the transformations will be
represented in terms of superdeterminants of certain super Wronski
matrices. This and other
related results will be presented elsewhere, \cite{lm}.

\newpage

\begin{center}
{\bf \Large List of captions}
\end{center}
\vspace*{2cm}

\begin{itemize}
\item[\bf Figure 1:] The function $f(x,t)$ plotted in the $x,t$ plane, is just
the standard 1-soliton of the KdV equation. 
\item[\bf Figure 2:] The function $g(x,t)$ plotted in the $x,t$ plane, it
represents a exponentially localized regular solution in $x$ that
travels with constant speed. 
\end{itemize}
\newpage

\vspace*{-4cm}
\leavevmode

\epsfxsize=14cm

\hspace*{0cm}\epsffile{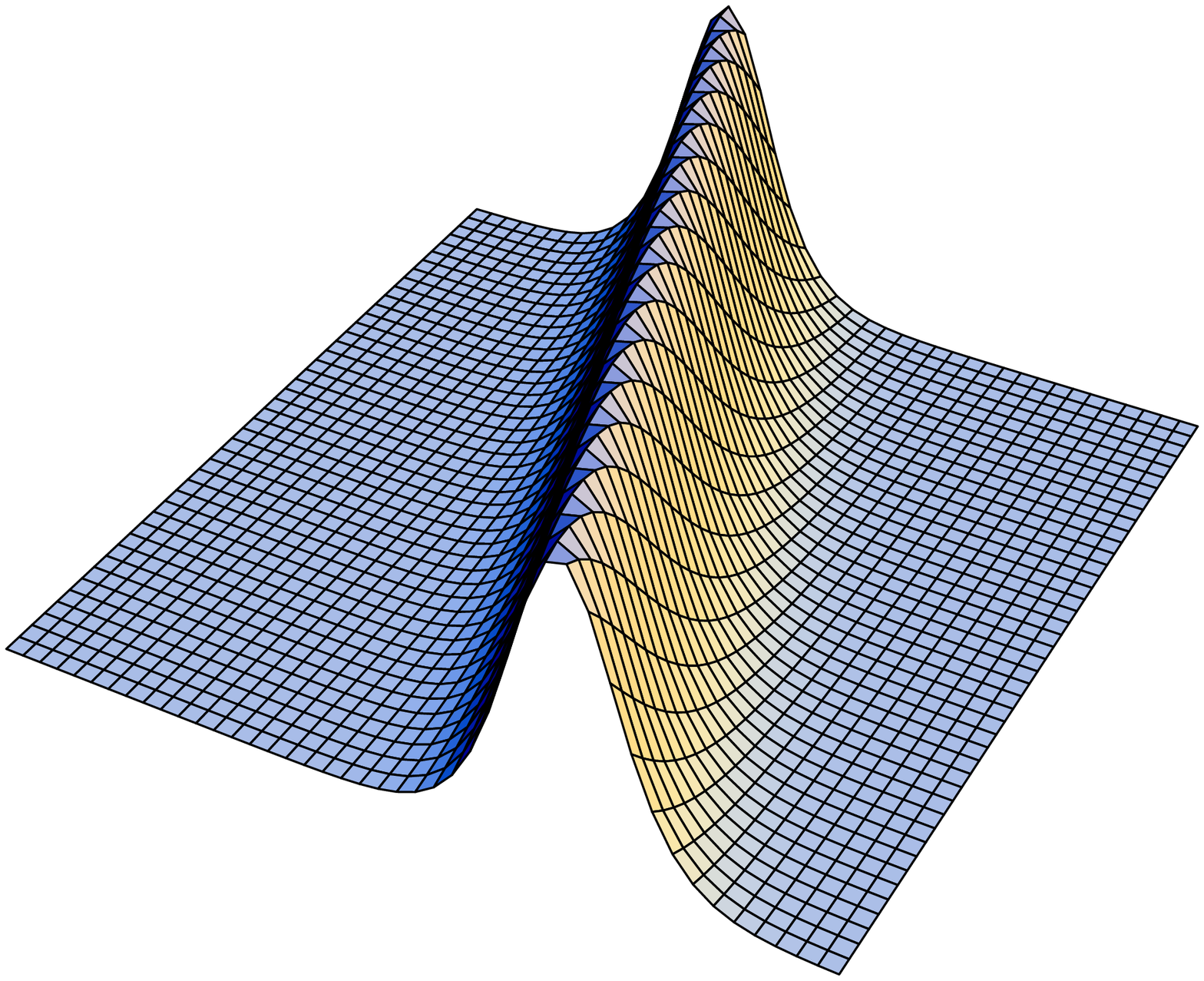}

\vspace*{-2cm}
\begin{center}
{\bf Figure 1}
\end{center}

\newpage

\vspace*{-4cm}
\leavevmode

\epsfxsize=14cm
\hspace*{0cm}\epsffile{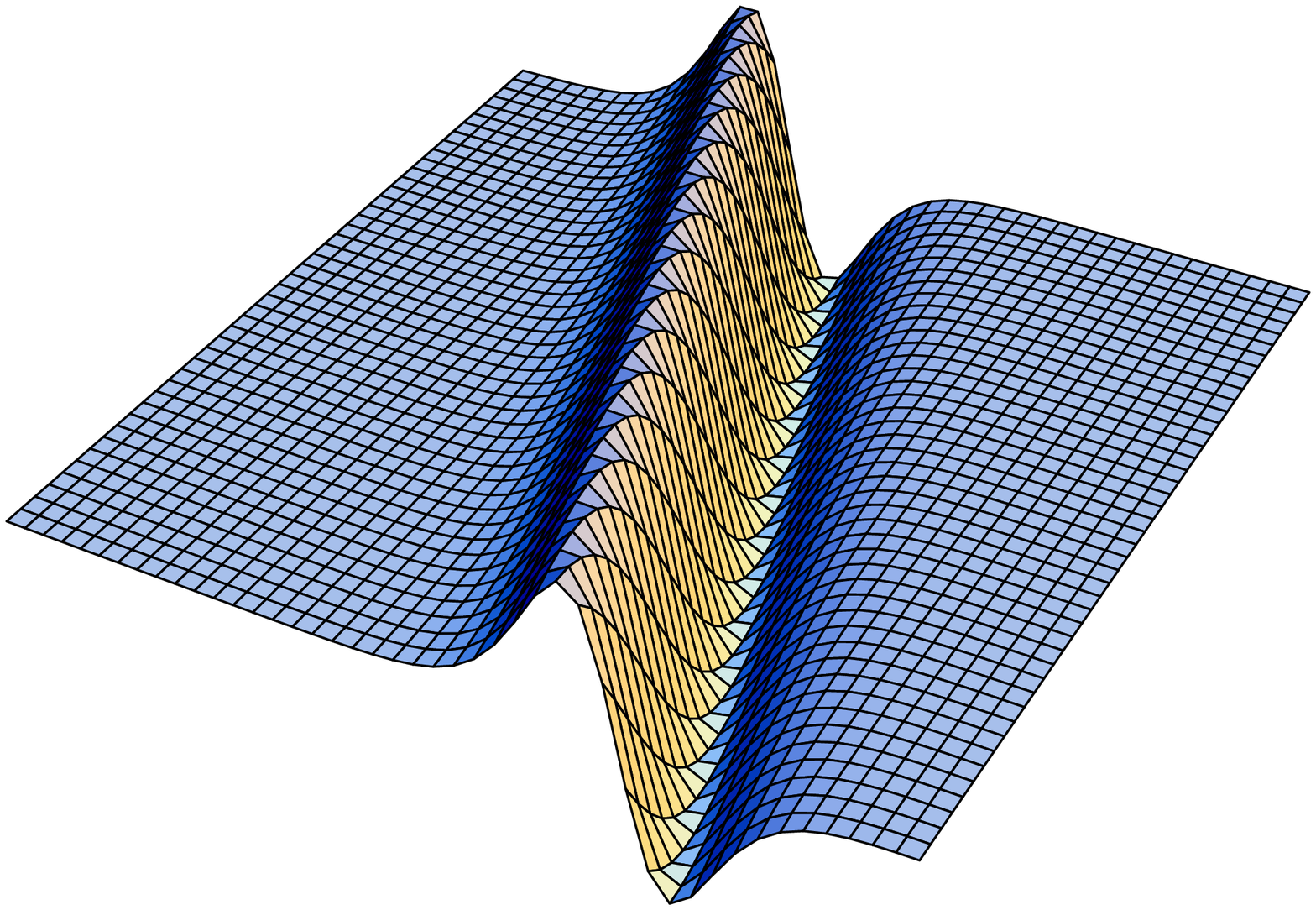}

\vspace*{-2cm}
\begin{center}
{\bf Figure 2}
\end{center}

\end{document}